\documentclass[superscriptaddress,prl]{revtex4}
\usepackage{epsfig,graphics}
\usepackage{graphicx}

\begin{document}
\title{Encapsulation of a fiber taper coupled microtoroid resonator in a polymer matrix}

\author{Faraz~Monifi,
        \c{S}ahin~Kaya~\"{O}zdemir,
         Jacob~Friedlein,~
        and~Lan~Yang,
\thanks{F. Monifi, \c{S}.K. Ozdemir, and L. Yang are with the Department
of Electrical and Systems Engineering, Washington University, St. Louis, MO, 63130 USA e-mail: (monifi@ese.wustl.edu, ozdemir@ese.wustl.edu, yang@ese.wustl.edu).}
\thanks{J.Friedlein is now with Department of Electrical, Computer, and Energy Engineering at the University of Colorado at Boulder, e-mail: (jacob.friedlein@ colorado.edu).}}

\begin{abstract}
We encapsulated a high-quality ($Q$) factor optical whispering gallery mode (WGM) microtoroid resonator together with its side coupled fiber taper inside a low refractive index polymer, achieving a final $Q$ higher than $10^7$. Packaging provides stable resonator-fiber taper coupling, long-term maintenance of high-$Q$, a protective layer against contaminants, and portability to microtoroid resonator based devices. We tested the robustness of the packaged device under various conditions and demonstrated its capability for thermal sensing.
\end{abstract}

\maketitle

\section{Introduction}

Ultra-high-$Q$ optical WGM resonators are currently under intensive investigation for their potential uses in photonic circuits \cite {1,2,3,4,5,6}, nonlinear optics \cite{7,8}, lasing \cite{9,10,11}, optical sensing \cite{12,13,14}, and optomechanics \cite{15,16}.
Despite their potentials to be the choice for many practical applications due to their ultra-high-$Q$, silica microsphere and microtoroid resonators have been lagging behind microring resonators. On-chip monolithic fabrication together with the coupling waveguides is challenging for silica microtoroid and microsphere resonators. Separately prepared optical fiber tapers are still the best means of coupling  ($>90\%$) light in and out of microspheres and microtoroids \cite{17}. However, maintaining coupling efficiency for long operation hours is difficult because the coupling conditions are open to environmental perturbations (e.g., air flow, mechanical vibrations, refractive index changes), requiring 3D nano-positioning systems and stabilization schemes.

When fiber-taper coupled microspheres or microtoroids are moved outside the laboratory settings, maintaining a stable coupling and high-$Q$ become even more difficult. For example, fiber tapers, which have diameters of the order of wavelength of the light used, are very fragile and need great care during the move and the operation in the field.  Moreover, equipping each resonator with a 3D nano-positioning system will not only make the system expensive and bulky, but also will require on-site or remote operator intervention. These limit the portability of devices based on fiber coupled microspheres and microtoroids. Finally, dust or other particulate matter and contaminants in uncontrolled environments may fall on the resonator and/or the fiber taper coupler, inducing extra scattering losses and perturbing the coupling condition. These significantly reduce $Q$ and make it very difficult, if not impossible, to maintain the same $Q$ and coupling conditions for long hours of operation.

A remedy to these obstacles is to isolate the fiber coupled microspheres and microtoroids from their surrounding by encapsulating them in suitable materials, which not only protects the device from environmental perturbations but also helps to maintain $Q$ and the coupling efficiency. Previously, Yan {\it et. al.} solved the aforementioned issues for fiber taper coupled microspheres by embedding either the whole device\cite{18} or only the coupling region in a low refractive index polymer\cite{19}. Although the former provides the desired isolation and stable coupling, it limits the sensing applications as the package minimizes, if not prevents, the interaction of the evanescent field with the surrounding outside the package. In the latter, on the other hand, the authors realized a spot-packaging which provides robust coupling efficiency without isolating the resonator completely from the environment so that it can still be effectively used for bio- or nanoparticle- sensing applications. Depending on the targeted application, one may prefer one over the other.

To date, a packaging strategy for fiber-coupled on-chip microtoroids has not been proposed or demonstrated. This is partly due to the fact that microtoroid packaging is extremely much more  challenging for the on-chip micro toroid. Below we have listed some of difficulties in packaging the microtoroid in comparison with microspheres.

Microtoroids usually possess one order of magnitude lower Q than microspheres therefore maintaining the gap between the taper and the microtoroid (maintaining the coupling condition to achieve the highest Q possible) is much more crucial in their packaging process. In the case of microspheres one can stabilize the taper by placing it in contact with the microsphere because even in such a coupling-loss dominated operating condition, it is still possible to obtain very high Q in the order of $10^7$ but this is not the case for microtoroid.

On the other hand, microtoroid resonators of major diameter equal to the diameter of microspheres have much smaller mode volumes. Because of the special geometry of microtoroid, optical mode is confined in the relatively small toroidal area and the azimuthal expansion of the mode is significantly smaller than that of microspheres. In case of microtoroid, the width of the effective coupling area is a limited area within the minor diameter of the toroid. Consequently small changes in height of the fiber-taper relative to microtoroid can completely quench the light coupling to the resonator. Note that almost in every position, regardless of how far the taper is from the equator of the microsphere, always one or more coupled resonance lines are observable. This is not the case for a microtoroid. Thus for that microsphere, both the mode volume and the effective coupling area are very large, which makes the packaging process much easier because the problems, such as narrower coupling regime, vertical and horizontal motions of the taper, difficulty in monitoring, related with the process are not very significant.
Also, since the microtoroid resonators are fabricated on a silicon wafer, the only way to monitor the packaging process is using microscopes which allows top and side views. As soon as the polymer is introduced, the curvature of the polymer droplet surface acts like a lens which distorts the image (i.e., in some cases one may completely lose the image and view) making the monitoring of the process very difficult.

  Here, we report encapsulation of an on-chip microtoroid resonator inside a low refractive index polymer together with its coupling fiber taper. We demonstrate that the packaged device is robust to environmental disturbances and maintains a high quality factor $Q\geq 10^7$.
\section{Design considerations}
A suitable encapsulation material should not significantly degrade the performance of the resonator from its performance in air (maintain $Q$ and mode volume $V$) yet provide the desired isolation from environment. This requires that it has a very low refractive index, and it is low-loss in the wavelength band of the light to be used. The former ensures a high refractive index contrast so that light is strongly confined within the resonator and only a small evanescent field leaks to the polymer (small $V$). This also leads to lower radiation losses for fixed resonator size. The latter, on the other hand, ensures that the evanescent field penetrating into the polymer region does not experience significant loss, keeping $Q$ as close to its value in air as possible.

In this study, we used the UV curable polymer MY $133$ (MY Polymers, Israel) which has refractive index of $1.331$ at $950 {\rm nm}$ wavelength after curing and very low loss for wavelengths higher than $450 {\rm nm}$. The polymer solution was provided as pre-filtered to below $0.5$ microns.

\begin{figure}
\includegraphics[width=3.4in]{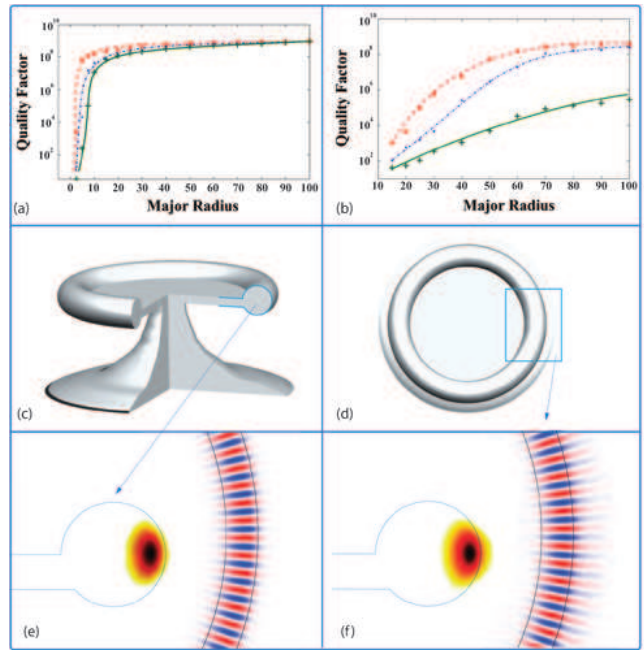}
\caption{Results of numerical simulations comparing quality factor $Q$ (A\&D), mode volumes (B\&E) and the mode distributions (C\&F) for unpackaged (A,B \& C) and packaged (D,E \& F) silica microtoroid resonators. Dependence of $Q$ on the size of the resonator is depicted at wavelength bands of $660 {\rm nm}$ (dashed), $980 {\rm nm}$ (dotted), and $1500 {\rm nm}$ (solid). For a silica toroid of major diameter $120 \mu {\rm m}$, mode volume is smaller for the unpackaged resonator (B) than for packaged resonator (E). Mode distribution profile shows that evanescent field penetrates more deeply into the surrounding for a packaged resonator than an unpackaged one. The packaging material is the polymer MY $133$. Simulations were carried out using COMSOL.}\label{Fig1}
\end{figure}

Numerical simulations imply that $Q$ is lower and $V$ is larger when a silica microtoroid resonator of a given size is encapsulated in MY$133$ (Fig. \ref{Fig1}). Larger $V$ is attributed to the lower refractive index contrast whereas lower $Q$ is attributed to the increased radiation loss resulting from the lower refractive index contrast and increased material loss due to additional loss experienced by the evanescent field penetrating into the polymer. Fig. \ref{Fig1} shows that, silica microtoroids larger than $40 \mu m$ in radius can acheive $Q>10^8$ in air for different wavelengths ($660{\rm nm}-1.5{\rm\mu m}$). This is attributed to the high refractive-index contrast which helps satisfy near-ideal total internal reflection. In polymer matrix, on the other hand, the contrast is low; hence it becomes difficult to satisfy the total internal reflection at the curved boundary of the microtoroid and the radiation (bending) loss increases as the wavelength increases for a given resonator size. It is seen in Fig. \ref{Fig1} that the quality factor increases with toroid size initially but then saturates at some diameter. This saturation diameter is greater at all wavelengths for toroids in polymer than for toroids in air. This implies that the toroid to be encapsulated in the polymer matrix should have a larger size to achieve the same $Q$ as the unpackaged resonator.

\begin{figure}
\centering
\includegraphics[width=3.5in]{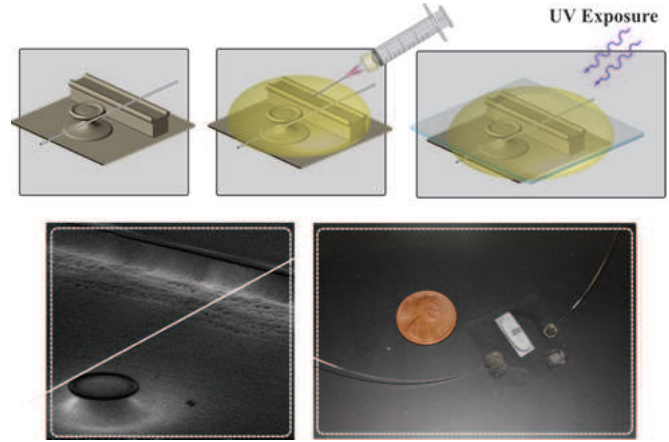}
\caption{Illustration of packaging process of fiber taper-coupled silica microtoroid resonator together with the reflowed silica side wall (upper panel), SEM image of a silica microtoroid and the reflowed silica wall (lower panel, left), and the optical microscope image of packaged silica microtoroid resonator (lower panel, right). Comparison with US penny coin is also given.}
\label{Fig2}
\end{figure}
\section{Experiments}
We fabricated microtoroids with major diameter $\sim 120 \mu m$, minor diameter $\sim 10\mu m$, and intrinsic Q factor $Q_0=5\times10^7$ according to the procedure given by Armani {\it et. al.} \cite{20} with a slight modification, which includes a reflowed silica wall beside each microtoroid on which the fiber taper waveguide could rest (Fig. \ref{Fig2}). These side walls limit the motion of the fiber taper due to the stress induced on it during the injection and the curing of the polymer solution. Thus, the vertical and horizontal coupling conditions between the fiber taper and the resonator are minimally affected during the packaging process. Moreover, since the side wall with very smooth surface has very small contact area with the fiber taper waveguide, the loss induced by side wall is minimal.

The chip containing the resonators was placed on a $3$D nanopositioning stage with $10$nm resolution. A fiber taper fabricated from standard single-mode fiber by heat-and-pull method over hydrogen flame was used as the waveguide to couple light into and out of the resonator. For vertical alignment of the fiber-resonator coupling, the stage was raised until the toroid was at the same height as the fiber taper suspended above the chip. In this position, the taper was in contact with the wall, and further increase of the stage does not change the relative heights of the taper and the resonator. Horizontal coupling was achieved by tuning the gap between the resonator and the fiber taper by horizontally moving the stage. Once this coupling is optimized, the toroid was raised further so that the taper is pressed more firmly against the wall. This prevents the taper from moving during later stages of packaging. In particular, the wall prevents the taper from moving up and down and reduces horizontal motion. During alignment process, the transmission spectra was continuously monitored by a photodetector connected to an oscilloscope.

\begin{figure}
\centering
\includegraphics[width=3.5in]{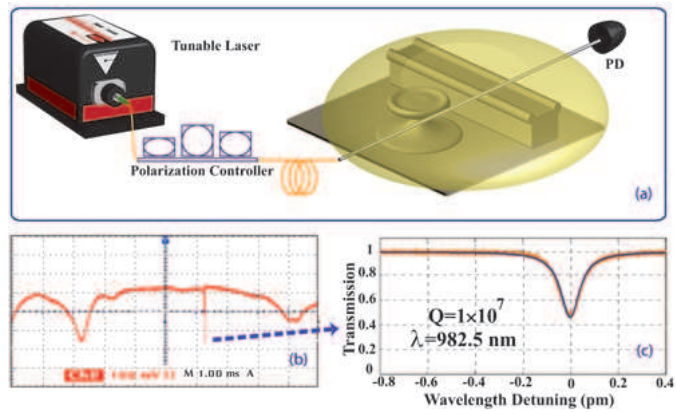}
\caption{(a) Illustration of the basic set-up for testing and measuring the transmission spectra of the fiber taper-coupled silica microtoroid and accompanying reflowed silica wall encapsulated in a low refractive index polymer. (b) Transmission spectra of the packaged microresonator when the wavelength was scanned within large span, and (c) enlarged spectra showing a high-$Q$ WGM resonance at $\lambda=982.5 \rm{nm}$ with $Q=1\times 10^7$.}
\label{Fig3}
\end{figure}

After the initial alignment, the polymer was applied using a syringe to cover the entire chip. A glass slide, which rests on spacers, was placed on top of the polymer layer to seal it from ambient oxygen since the curing takes place in the absence of oxygen. During the process, horizontal alignment was monitored and adjusted as needed. Upon achieving a slightly undercoupled condition, we applied UV light for about 10 minutes to cure the polymer. The curing process shrinks the polymer and induces extra stress on the fiber taper, pulling it towards the microtoroid and hence bringing the system close to critical coupling region. Finally, the rest of the taper (the parts that are not directly above the chip) is coated with the same polymer to completely isolate the system from the surrounding. The whole packaging process is depicted in Fig. \ref{Fig2}. We used an external-cavity, tunable laser in the $980 {\rm nm}$ wavelength band to probe the resonances and transmission spectra of the microtoroids in air and in the polymer matrix (Fig. \ref{Fig3}). For toroids with $Q$ measured as $5\times 10^7$ in air, we obtained $Q\sim 10^7$ and $Q\sim 6.8\times10^6$ when they were packaged to be in the deep undercouping and critical coupling regions, respectively. Note that in the deep undercoupling regime, intrinsic losses dominates the coupling losses and the measured $Q$ is approximately equal to the intrinsic quality factor, $Q_0$. These results clearly show that the performed packaging alters the $Q$ only slightly.

\subsection{Application in Aquatic Environment}

\begin{figure}
\centering
\includegraphics[width=3.5in]{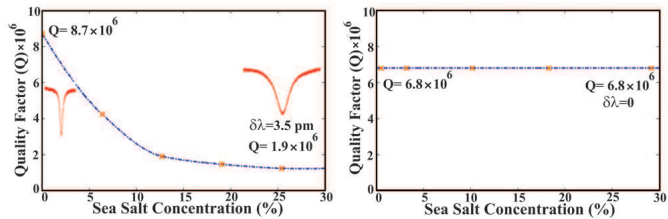}
\caption{Comparison of the change in the $Q$ and WGM resonance wavelengths for unpackaged (left) and packaged (right) silica microtoroids placed in water environment with varying salt concentrations. Packaged resonator maintains its resonance wavelength and $Q$ for all salt concentrations.}
\label{Fig4}
\end{figure}

Next, we assess the quality of the packaging process in isolating the microtoroid resonator from the surrounding by placing it in water with varying salt concentrations. As it is seen in Fig. \ref{Fig4}, increasing salt concentration in the water environment significantly degrades the $Q$ of the unpackaged microtoroid whereas it does not affect $Q$ of the microtoroid encapsulated in the MY133 polymer matrix. Unpackaged microtoroid experiences $\sim 10$-fold decreases in its $Q$, in addition to $3.5 {\rm pm}$ of resonance shift, when the salt concentration is increased from $0\%$ to $25\%$ . The packaged microtoroid, on the other hand, experiences neither $Q$ degradation nor resonance shift, implying that the evanescent tail of the WGM field of the microtoroid is completely contained within the polymer package with no leakage into the water surrounding.

\subsection{Thermal sensing}
\begin{figure}[h!]
\centering
\includegraphics[width=3.5in]{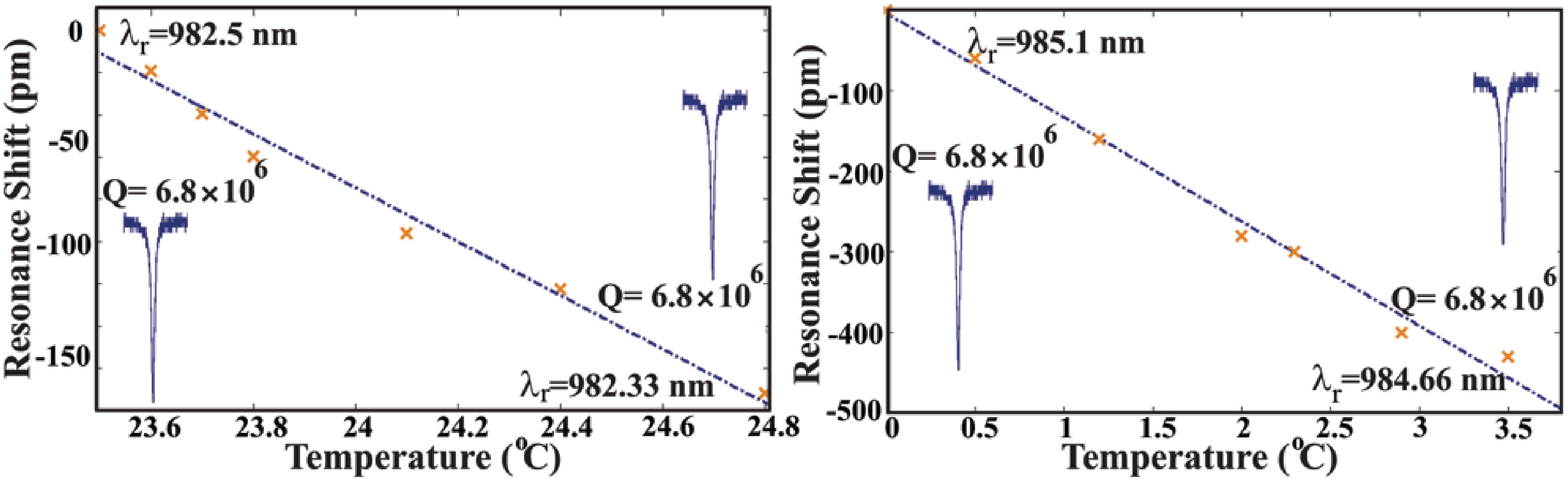}
\caption{Thermal sensing performance of packaged silica microtoroid resonator placed in air with temperature controlled using a thermo-electric cooler (TEC) (left) and in cold water with temperature controlled by gradually introducing ice into the water chamber (right). Thermal sensitivity is the same in both cases.}
\label{Fig5}
\end{figure}

We tested the possibility of using this packaged structure for thermal sensing in air and in water. We placed the device on a thermo-electric cooler (TEC) to increase the temperature. We found that for a $1 ^\circ C$ increase in temperature, the resonant frequency increased by $60 {\rm GHz}$, implying that the packaged toroid is highly sensitive to changes in the temperature of its surrounding. The temperature sensitivity (resonance wavelength shift per unit temperature change) of the packaged microtoroid resonator was determined as  $0.131 nm/ ^\circ C$ by linear fitting to the experimentally observed resonance shift as a function of temperature (Fig. \ref{Fig5}). Similar experiment with unpackaged silica microtoroid resonator demonstrated a temperature sensitivity of $0.0117 nm/ ^\circ C$, which is $\sim 1/10$-th of that for the packaged microtoroid. It is to be noted here that in contrast to an unpackaged silica microtoroid, the packaged one experiences blue shift with increasing temperature, implying that thermo-optic coefficients of MY-$133$ and silica have opposite signs, i.e.,negative for MY-$133$ and positive for silica.

Next, we placed the packaged microtoroid resonator into cold-water environment and by gradually adding ice we decreased the temperature of water down to $0.5-4 ^\circ C$ range. For the temperatures in this range, we took the transmission spectra after the homogenization of the temperature in the chamber. The linear relation between temperature and the resonance shift with the same thermal sensitivity of  $0.131 nm/ ^\circ C$ as the test in air with TEC was observed (Fig. \ref{Fig5}), suggesting that the packaged resonator is equally sensitive to changes in high and low temperatures.

\section{Conclusion}
In summary we have embedded a silica microtoroid resonator side coupled to a tapered fiber into a low refractive index polymer, which provides both mechanical and optical stability, i.e., long term stable coupling and retained high-$Q$ in different environments. We demonstrated $Q$ values higher than $10^7$ for packaged silica microtoroids, tested them in different surroundings and used them as temperature sensors. Packaging eliminates the need for stabilization techniques and 3D nanopositioning systems when the resonator is moved outside laboratory, making the field operation of devices based on WGM microtoroid and microsphere resonators cost effective. We believe that packaging will enable the use of WGM resonators in harsh environments such as undersea and deep-ocean and airborne crafts. One may apply over the low index polymer coating another protection layer made of a material that is robust against expected perturbations in the targeted harsh environment.


\begin{thebibliography}{1}

\bibitem{1}
F. Monifi, J. Friedlein, \c{S}. K. \"{O}zdemir and L. Yang, "A Robust and Tunable Add–Drop Filter Using Whispering Gallery Mode Microtoroid Resonator," \emph{J. Lightw. Technol.}, vol. 30, no. 21, pp. 3306-3315, Nov. 2012.
\bibitem{2}
Y. A. Vlasov, W. M. J. Green and F. Xia,"High-throughput silicon nanophotonic wavelength-insensitive switch for on-chip optical networks" \emph{Nat Photonics}, vol. 2, pp. 242-246, Mar. 2008.

  \bibitem{3}
 F. C. Blom, D. R. van Dijk, H. J. W. M. Hoekstra, A. Driessen and Th. J. A. Popma, "Experimental study of integrated-optics microcavity resonators: Toward an all-optical switching device" \emph{Appl. Phys. Lett.}, vol. 71, pp. 747-749 Aug. 1997.

  \bibitem{4}
 H. C. Tapalian, J. P. Laine and P. A. Lane, "Thermo-optical switches using coated micro-sphere resonators" \emph{IEEE Photon. Technol. Lett.}, Vol.14, pp. 1118-1120 Aug. 2002.

  \bibitem{5}
 M. Lipson, "Compact Electro-Optic Modulators on a Silicon Chip" \emph{IEEE J. Sel. Top. Quantum Electron.} Vol. 12, no. 6, pp. 1520-1526, Nov. 2006.

  \bibitem{6}
L. Maleki, A. B. Matsko, A. A. Savchenkov and V. S. Ilchenko, "Tunable delay line with interacting whispering-gallery-mode resonators" \emph{Opt. Lett.}, Vol. 29, no. 6, Mar. 2004.

\bibitem{7}

Y. F. Xiao, S. K. Ozdemir, V. Goddam, C. H. Dong, N. Imoto and L. Yang, "Quantum nondemolition measurement of photon number via optical Kerr effect in an ultra-high-Q microtoroid cavity", \emph{Opt. Express }, Vol. 16, no. 26 ,pp. 21462-21475, Dec. 2008.

  \bibitem{8}
A. B. Matsko, A. A. Savchenkov, D. Strekalov, V. S. Ilchenko, and L. Maleki, "Review of Applications of Whispering-Gallery Mode Resonators in Photonics and Nonlinear Optics", \emph{IPN Progeress Report} Vol.42-152, Aug. 2005.


  \bibitem{9}
  L. He, S. K. Ozdemir, J. Zhu and L. Yang, "Self-pulsation in fiber-coupled, on-chip microcavity lasers", \emph{Opt. Letters} Vol. 35, no. 2, pp. 256-258, Jan. 2010.

  \bibitem{10}
L. Yang, T. Carmon, B. K. Min, S. M. Spillane, and K. J. Vahala, "Erbium-doped and Raman microlasers on a silicon chip fabricated by the sol–gel process", \emph{Appl. Phys. Lett.}, vol. 86, no. 9, pp. 091114, Feb. 2005.

  \bibitem{11}

S. M. Spillane, T. J. Kippenberg and K. J. Vahala, "Ultralow-threshold Raman laser using a spherical dielectric microcavity", \emph{Nature} Vol. 415, pp. 621-623, Feb. 2002.

  \bibitem{12}

L. He, S. K. Ozdemir, J. Zhu, W. Kim and L. Yang, "Detecting single viruses and nanoparticles using whispering gallery microlasers", \emph{Nat Nanotechnol.}, Vol. 6, pp. 428-432, Jun. 2011.

  \bibitem{13}

J. Zhu, S. K. Ozdemir, Y. F. Xiao, L. Li, L. He, D. R. Chen and L. Yang, "On-chip single nanoparticle detection and sizing by mode splitting in an ultrahigh-Q microresonator" \emph{Nat Photonics},Vol. 4, pp. 46-49, Dec. 2009.

  \bibitem{14}

X. D. Fan, I. M. White, S. I. Shopova, H. Y. Zhu, J. D. Suter and Y. Z. Sun, "Sensitive optical biosensors for unlabeled targets: A review", \emph{Anal. Chim. Acta} Vol. 620, no. 1-2, pp. 8-26, May 2008.

  \bibitem{15}

  T. J. Kippenberg and K. J. Vahala, Science, "Cavity Optomechanics: Back-Action at the Mesoscale" \emph{Science}, Vol. 321, no. 5893 pp. 1172-1176, Aug. 2008.

  \bibitem{16}

G. S. Wiederhecker, S. Manipatruni, S. Lee and M. Lipson, "Broadband tuning of optomechanical cavities", \emph{Opt. Express}, Vol. 19, no.3, pp. 2782-2790, Jan. 2011.


  \bibitem{17}

M. Cai, O. Painter and K. J. Vahala, "Observation of Critical Coupling in a Fiber Taper to a Silica-Microsphere Whispering-Gallery Mode System", \emph{Phys. Rev. Lett.}, Vol. 85, pp. 74–77, Jul. 2000.

 \bibitem{18}
Y.-Z. Yan, {\it et. al.},"Packaged silica microsphere-taper coupling system for robust thermal sensing application"  \emph{Opt. Express} , Vol. 19, no. 7, pp. 5753-5759, Mar. 2011.

\bibitem{19}
Y.-Z. Yan, {\it et. al.},"Robust Spot-Packaged Microsphere-Taper Coupling Structure for In-Line Optical Sensors" \emph{IEEE Photon. Technol. Lett.} , Vol. 23 , no. 22, pp. 1736-1738, Nov. 2011.

\bibitem{20}
D. K. Armani, T. J. Kippenberg, S. M. Spillane and K. J. Vahala,"Ultra-high-Q toroid microcavity on a chip", \emph{Nature}, Vol. 421, pp. 925-928, Feb. 2003.

\end{thebibliography}
\end{document}